# Electronic interactions in strongly correlated systems: what is the glue for high temperature superconductivity?

Tonica Valla[*]

Department of Physics, Brookhaven National Laboratory, Upton, NY 11973-5000, USA

## ABSTRACT

Recent observation of a "kink" in single-particle dispersion in photoemission experiments on cuprate superconductors has initiated a heated debate over the issue of a boson that "mediates" the pairing in cuprates. If the "kink" is indeed caused by interaction with a bosonic excitation, then there are two possible candidates: phonons and spin fluctuations. Here, the role of anti-ferromagnetic spin fluctuations in shaping the phase diagram of cuprate superconductors will be discussed. By using the local (momentum-integrated) dynamic spin susceptibility, recently measured in neutron scattering experiments to high energies, the electronic self-energies are calculated that agree in many aspects with those measured directly in angle-resolved photoemission (ARPES) and optical spectroscopies. The spin fluctuations therefore seem to play a role typically played by phonons in renormalizing single particles. The key question emerging from this picture is whether the coupling detected in ARPES reflects the mediating boson, i.e. whether the spin fluctuations may be responsible for superconducting pairing.

## 1. INTRODUCTION

Recently, the high quality photoemission data on cuprate superconductors uncovered the "kink" in single-particle dispersion – an indication of coupling between the electrons and some bosonic excitation.[1,2,3,4,5,6,7,8] This discovery has re-initiated speculations about the origin of high temperature superconductivity (HTSC) due to a possibility that the observed coupling reflects the boson involved in pairing. The optical conductivity also indicates the existence of similar bosonic excitations coupled to carriers. Different methods have been developed to extract the bosonic spectrum from optical scattering rates[9,10,11] and soon the contours of the spectrum started to emerge, resembling those extracted from ARPES: a mode or high density of boson states at 50-80 meV and a broad continuum extending to very high energies must both be present in the spectrum. However, even with more details becoming available, neither probe has been able to answer the key question what the mode is. Two main candidates have attracted most attention: a) phonons and b) the so called spin resonance.

The most direct effects of electron-phonon coupling ("kinks") in conventional metals have became visible in ARPES only recently,[12,13,14] slightly before the "kinks" in cuprates were detected.[1] An example is shown in Fig. 1(a). Due to similarities with effects in cuprates (see Fig. 2), the conventional electron-phonon coupling[15,16] (see for example the Ashroft/Mermin's[17] and Mahan's[18] textbooks) has been initially seen as an exclusive mechanism that produces "kinks" and which is at play again in cuprates.[4] However, it has been realized soon that the same effects may be produced by charge-density wave gaps[19,20] and, in conventional magnets, by magnetic fluctuations (magnons),[21] as shown in Fig. 1(b) and 1(c).

There are several problems if one wants to attribute the observed single particle self-energies and optical conductivities in the cuprate superconductors to conventional electron phonon coupling. Firstly, the ever increasing $Im\Sigma(\omega)$ indicates that the spectrum of excitations ranges to at least 200-300 meV as the scattering rate shows no signs of saturation even at these energies.[1,3,6,9,10,11] Phonons are limited to 80-90 meV or less.[22,23,24,25] Secondly, the kink shows a significant doping dependence, monotonically decreasing with doping[4,5,7,8] with some evidence that it essentially disappears at $x > 0.3$, when superconductivity also vanishes.[6] Phonons, however, are still present for $x > 0.3$. Thirdly, the kink,[4,5,7,8] as well as the low energy scattering rates[3,9,10] and consequently the photoemission line-shape,[26,27] are temperature dependent with a pronounced change at $\sim T_C$ in systems with a high $T_C$ near optimal doping. These latter two facts would suggest that strong changes in the coupling and in the phonon spectrum itself would have to occur with doping and temperature. This has not been observed.

---

[*] Further author information:
E-mail: valla@bnl.gov, Telephone: 1 631 344 3530

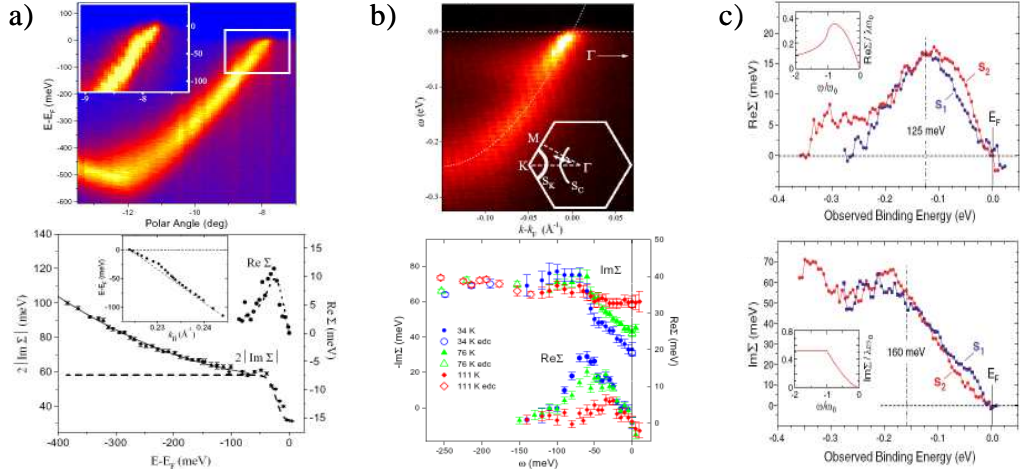

**Figure 1:** Photoemission spectra and the corresponding self-energies for several conventional systems: **a)** Mo(110) surface state,[12] a conventional 2-D metal, **b)** 2H-TaSe2, a layered CDW material[19] and **c)** Fe(110) surface state, a prototypical ferromagnet.[21] "Kinks" exist in all three systems. In a) the self energy is due to the electron-phonon interaction. In b) and c) the energy scale is several times higher than the corresponding phonon cut-off and the electron-phonon coupling is not a viable origin of the kinks.

Another frequently discussed candidate for the coupling measured in both ARPES and optical conductivity is spin fluctuations. Often, the commensurate $(\pi,\pi)$ "resonance"[28,29,30,31] is considered as the sole cause of the observed coupling effects.[26,27,32,33,34] This picture, however, suffers from similar problems as the phonon scenario. If the "resonance" alone causes the coupling, then Im$\Sigma(\omega)$ should saturate above the resonance energy $\Omega_{res}$. The temperature and doping dependences would fit qualitatively better into the "resonance" scenario as the quasi-particle (QP) coupling decreases with doping and the self energy mimics the temperature behavior of the resonance mode. There is, however, one additional problem: "kinks" also exist in systems as $La_{2-x}Sr_xO_4$, where the "resonance" mode had not been detected.

## 2. SPIN FLUCTUATIONS AND ELECTRONIC SELF-ENERGIES

In this paper, we will point to several experimental facts that suggest very close relationship between spin fluctuations and single-particle properties and superconductivity itself. We will demonstrate that most of the controversial aspects of ARPES and optical conductivity may be accounted for, if, instead of the "resonance mode", the whole spin-fluctuation spectrum is used to determine the QP self-energies. For this purpose, the high-energy (up to 300 meV) dynamic spin susceptibility data are necessary. Recently, neutron scattering experiments have been performed on $La_{2-x}Ba_xO_4$,[35] a system very similar to $La_{2-x}Sr_xO_4$ and on underdoped YBCO,[36,37] up to very high energies (~300 meV). These experiments have shown that the spin excitations are remarkably similar among different families of cuprate superconductors, consisting of a commensurate $(\pi,\pi)$ scattering at some finite energy $\Omega_{res}$, and scattering branches dispersing downwards and upwards out of this $(\pi,\pi)$ mode with increasing incommensurability. For $La_{2-x}Ba_xO_4$, at $x=1/8$, the low energy branch goes all the way to zero energy and the system is in a disordered magnetic ("static stripe") phase, with a very suppressed superconductivity: $T_C$ is lowered below 4 K. Here, we will model QP self energies by using the $La_{2-x}Ba_xO_4$ susceptibilities and compare them to those that have been already measured in ARPES[6] on $La_{2-x}Sr_xO_4$. We will also model the self-energies for systems with much higher $T_C$ and with so called "spin resonance" mode. The low energy ($\leq 50$meV) spin susceptibility in these systems is markedly temperature dependent. In the superconducting (SC) state, the spin spectrum is gapped and relatively well-defined modes and a strong commensurate "resonance" exist, while in the normal state (NS), (or above the pseudogap temperature, in the underdoped systems[38]), the spin gap closes, and the excitations get overdamped and lose identity. Due to these changes, the calculated QP self-energies are also strongly temperature dependent, in agreement with the experiments.

## 2.1. Kinks and scattering rates

As the first exercise, the local (momentum integrated) spin susceptibility from ref. 35 is simply put into formulae for QP self-energies instead of the usual phonon $\alpha^2 F$. The resulting self-energy reproduces virtually all the essential attributes measured in ARPES. Fig. 2(c), reproduced from ref. 35, shows the local susceptibility $\chi''(\omega)$, or effectively, the density of states (DOS) of bosonic excitations in $La_{2-x}Ba_xO_4$ at $x = 0.125$. Aside from the incommensurate low energy structure related to the spin ordering, $\chi''(\mathbf{q},\omega)$ has a commensurate peak at ~55 meV that disperses in the same way as the "resonance" mode in YBCO.[36,37] Also, the magnetic spectrum clearly extends to ~250 meV. If $\chi''(\omega)$ is simply taken as $\alpha^2 F$ - bosonic DOS spectrum, the electronic self-energies may be calculated using

$$\mathrm{Im}\Sigma(\omega,T) = \pi \int_0^\infty d\nu \alpha^2(\nu) F(\nu) [2n(\nu,T) + f(\nu+\omega,T) + f(\nu-\omega,T)], \quad (1)$$

where $\alpha^2 F(\omega)$ is the Eliashberg coupling function, and $f(\omega,T)$ and $n(\omega,T)$ are the Fermi and Bose-Einstein distributions. Clearly, in the case of an anisotropic system, equation (1) is an oversimplification as the electronic states at different momenta may couple to different bosonic states with different strengths, resulting in $k$-dependent self-energy. A proper picture would involve $k$-resolved calculations for any initial state coupled to all possible final states trough full $\chi''(\mathbf{q},\omega)$. A $k$-dependent (pseudo)gap will complicate the situation further. At the least, (pseudo)gaps will affect the DOS of initial/final electronic states involved in the scattering.[39] However, even a poor approximation as eq. (1), gives surprisingly good results that qualitatively agree with measured self-energies. Obviously, Im$\Sigma$ monotonically increases with energy over the region over which the bosonic spectrum exists. Re$\Sigma(\omega)$ may be obtained from Im$\Sigma(\omega)$ by using Kramers-Kronig relations. The resulting $\Sigma(\omega)$ is shown in Fig. 2(b) and (d).

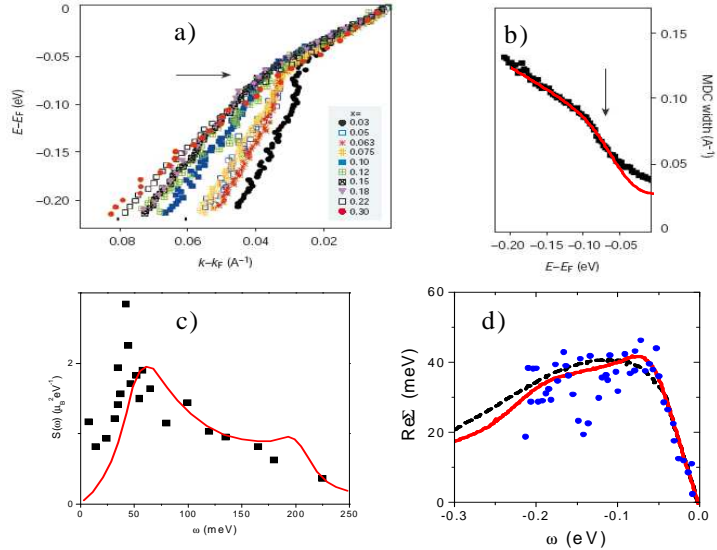

**Figure 2**: **a)** Nodal line dispersions for $La_{2-x}Sr_xO_4$ for $0.03 \leq x \leq 0.3$ at 20 K reproduced from ref. 6 and **b)** momentum width ($\propto$Im$\Sigma$) for $x=0.063$. Red line in b) is Im$\Sigma(\omega)$ calculated by using model $\chi''(\omega)$ from panel (c). **c)** Local spin susceptibility $\chi''(\omega)$ for $La_{2-x}Ba_xO_4$ ($x=0.125$) reproduced from ref. 35. The experimental points (squares) and the model spin density of states for spin-ladders (line) are shown. **d)** Re$\Sigma(\omega)$ extracted from dispersion in panel (a) for $x=0.12$ (points) and calculated by using experimental (dashed line) and model (solid line) $\chi''(\omega)$ from panel (c).

Re$\Sigma(\omega)$ and Im$\Sigma(\omega)$ in $La_{2-x}Sr_xO_4$ have been measured by Zhou *et al* [6] from $x=0.03$ to $x=0.3$ and are reproduced in Fig. 2a), and 2b). The kink in dispersion and the suppression in the scattering rate at $\approx 60$ meV are clearly visible. At higher energies, Im $\Sigma(\omega)$ is $\propto \omega$ with no signs of saturation. In Fig. 2(d), the calculated self-energy (real part) is shown and compared with the measured one for $x=0.12$. It is obvious that the typical choice for the non interacting dispersion that forces $Re\Sigma(\omega)$ to zero at $\omega \sim 200-250$ meV is not appropriate when the excitation spectrum extends to these energies. Therefore, we have chosen the bare dispersion which gives $Re\Sigma(\omega)$ that overlaps with the calculated one at the highest energy. The second direct evidence that the excitation spectrum extends that high is the $Im\Sigma(\omega)$ (Fig. 2(b)). Since eq. (1) is an integral over the excitation spectrum, the very fact that it does not saturate, points to the lack of cut-off in the excitation spectrum. If the local spin susceptibility is used as an effective $\alpha^2 F(\omega)$, all the features are reproduced, including the nearly linear increase in $Im\Sigma(\omega)$ at high energies and a reduction below $\omega \sim 60-70$ meV. It is instructive that once the susceptibility scale is adjusted to reproduce the magnitude of $Re\Sigma(\omega)$, the $Im\Sigma(\omega)$ is scaled properly, indicating that the self-energy concept still makes sense. As there is no dramatic change in the dynamic spin susceptibility with temperature in the 214 (LSCO and LBCO) class of materials, there can not be significant change in the QP self energy with temperature in these materials. In $Bi_2Sr_2CaCu_2O_{8+\delta}$, the system with big changes between the normal state and superconducting state susceptibilities,[31] significant changes in ARPES have also been observed at $T_C$. The kink and scattering rates change at $T_C$ near the nodal line,[3,5] and even more strongly further away, near the antinodes.[3,8] There, the self energies change quite dramatically, making an obvious change in the lineshape at $T_C$. The stronger coupling near the antinodes also points to spin fluctuations because the magnetic scattering concentrated to $(\pi,\pi)$ (antiferromagnetic) vector spans the antinodes better than the nodes. To model the self-energies for systems with the resonance mode and $T$-dependent susceptibility, we have used $\chi''(\omega)$ measured on YBCO from ref. 38 and compare the results with self-energies measured in $Bi_2Sr_2CaCu_2O_{8+\delta}$ at similar doping levels and temperatures. Results are shown in Fig. 3.

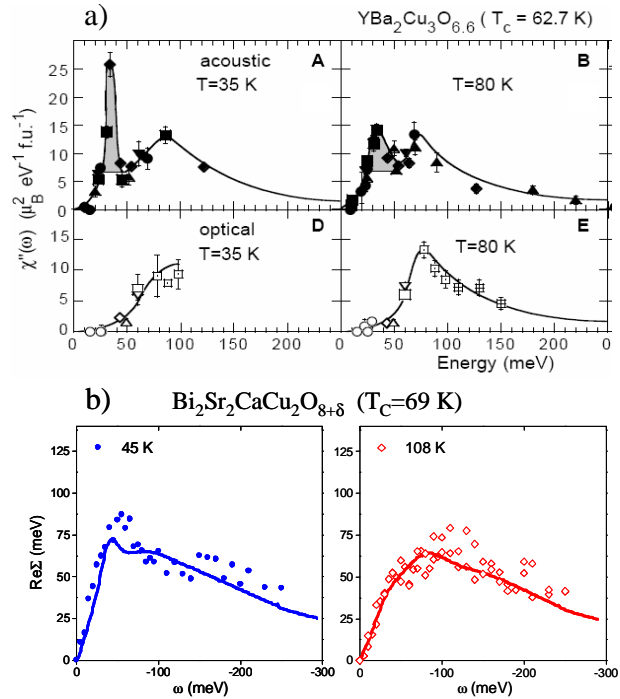

**Figure 3**: **a)** Local (momentum-integrated) dynamic spin susceptibility $\chi''(\omega)$ from ref. 38, for acoustic and optical modes in an underdoped YBCO below (left) and above (right) $T_C$. **b)** $Re\Sigma(\omega)$ for an underdoped Bi2212 measured in ARPES[5] (points) and calculated from acoustic YBCO susceptibilities from a) (lines).

Based on the doping dependence of the self energy measured in ARPES, it is possible to predict some global characteristics of $\chi''(\omega)$ as the doping is varied. The energy of the main peak and the overall energy range would probably not change dramatically with doping. However, the main peak should gain the relative weight as doping is lowered. Therefore, one could predict the magnetic scattering intensities for different doping levels, relative to those measured at 1/8 doping. For example, already for $x = 0.1$, the main peak in $\chi"(\omega)$ should be ~ 2 times more intense than that measured at 1/8 doping to produce the "kink" measured in ARPES. In the framework of the original model of Tranquada *et al*,[35] $x=0.1$ would correspond to the 3-leg ladder system, and $x = 1/12$ to the 4-leg ladder. As odd leg ladders are gapless, and even leg ladders have a spin gap,[40] one would expect an alternating behavior for the spin gap and the main peak structure originating from it, as the doping decreases. However, on going down from $x = 1/8$, a lack of static order, a coupling between the ladders and the existence of multiple modes in ladders with more legs would certainly complicate the picture. On the other hand, a simpler picture where susceptibility follows from the Fermi surface gapped with a *d*-wave gap, seems to be able to produce all the important attributes of dynamic susceptibility, including the measured dispersion, rotation of scattering maxima and doping dependence of scattering intensity.[32,41,42,43,44,45,46,47]

## 2.2. Overdoped side

The most important consequences, however, follow from the single particle spectra in the highly overdoped regime of the phase diagram as illustrated in Fig. 4. Note that the dispersion for $x = 0.3$ sample is essentially a straight line - there is almost no kink![6] The coupling to whatever boson there was at lower doping levels, nearly vanishes at 30% doping. A continuous weakening of coupling effects with increased doping (see Fig. 2a)) culminates in total vanishing at the same doping level where superconductivity disappears. The most obvious way to reduce the coupling is to reduce the bosonic DOS. If spin fluctuations is the boson that renormalizes ARPES spectra, then one could anticipate that $\chi"(\omega) \to 0$ for 30% doping. Recent work on highly overdoped $La_{2-x}Sr_xO_4$ samples shows that at least the low energy part of $\chi"(\omega)$ indeed vanishes at $x=0.3$, simultaneously with superconductivity.[48] If, as in conventional superconductors, the single particle dispersion signals the boson that causes pairing, the identity of that boson in the cuprates becomes now more suggestive. Put everything together, the following line of reasoning seems unavoidable: The coupling seen in ARPES is coupling to $\chi"(\mathbf{q},\omega)$ and is intimately related to superconductivity. As the boson DOS vanishes, coupling weakens and superconductivity eventually disappears. Note that unlike phonons, spin fluctuations really tend to disappear at 30% doping. A crucial experiment would be to extend the neutron scattering measurements to higher energies for $x \geq 0.3$ samples to test if susceptibility at higher energies also vanishes. Even though the low energy part of spin susceptibility is probably more closely related to superconductivity, it is the higher energy part that influences the single particle properties measured in ARPES more visibly. Therefore, based on a vanishing coupling in ARPES, one may anticipate diminishing of $\chi"(\omega)$ over the whole energy range.

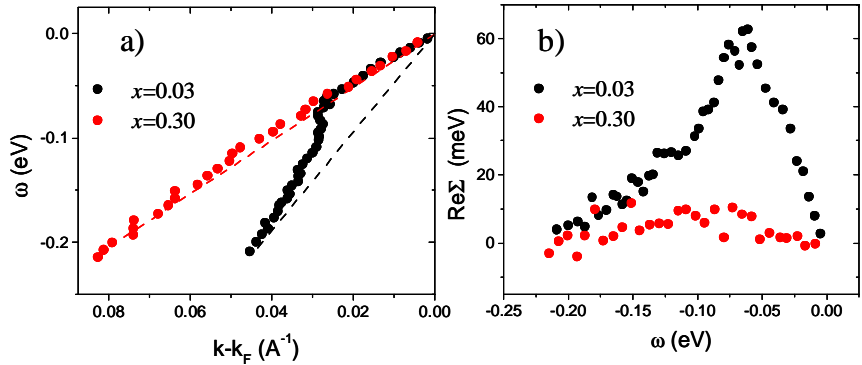

**Figure 4**: **a)** Nodal dispersions for $La_{2-x}Sr_xO_4$ from ref. 6 for two non-superconducting samples on opposite sides of the superconducting dome: $x=0.03$ (black circles) and $x=0.3$ (red circles). **b)** Re$\Sigma$ obtained from (a) by using the straight lines for non-interacting dispersions. Note nearly vanishing coupling for $x=0.3$.

## 2.3. Onset of superconductivity ($x \approx 0.05$)

There is another important evidence for a close relationship between superconductivity and spin fluctuations on the opposite side of the "superconducting dome" that emerges form ARPES as well as transport and optical data from the low-doping regime where superconductivity just appears. At $x = 0.055$ a peculiar transition has been observed in neutron scattering experiments: a static incommensurate scattering near AF wave vector with incommensurability $\delta \propto x$ rotates by 45°, from being diagonal to Cu-O bond ($x \leq 0.055$) to being parallel to it ($x \geq 0.055$).[49,50] This rotation coincides with the transition seen in transport[51] where "insulator" turns into superconductor for $x > 0.055$. Although the in-plane transport becomes "metallic" at high temperatures even at 1% doping, at low temperature, there is an upturn in $\rho_{ab}$, indicating some localization of carriers. This upturn is only present for $x < 0.055$, where the diagonal, static incommensurability exists. When diagonal points disappear (rotate) upon doping, the upturn in resistivity vanishes, and superconductivity appears instead. Now, recent ARPES experiments[52] have shown that the first states appearing at the Fermi level are those near the nodal point, whereas the rest of the Fermi surface is affected by a large gap of similar $d$-wave symmetry as the superconducting gap. While it is clear that the near-nodal states are responsible for metallic normal state transport at these low doping levels, their role in superconductivity is not adequately appreciated. We argue here that they are crucial for the onset of superconductivity. A closer look at ARPES results from this region of the phase diagram uncovers[52,53] that the increase in $k_F$ of nodal states is closely related to the diagonal incommensurability, $\delta$, seen in neutron scattering. Moreover, it appears that $\delta \approx \Delta k_F$, where $\Delta k_F$ is a nodal Fermi wave-vector measured from $(\pi/2,\pi/2)$, as can be seen from Fig. 5. This means that the diagonal elastic incommensurate scattering may be understood in terms of conventional spin-density wave (SDW) picture, originating from $2k_F$ nesting of nodal states. Such SDW would open the gap and localize the nodal states at low temperatures, preventing the superconductivity from occurring, in agreement with transport[51,54] and optical[55] studies. Note that the nodal $k_F$ increases proportionally to $x$ only for $x < 0.07$ while the diagonal incommensurability exists. Further doping destroys the diagonal SDW, releasing the nodal states. Superconductivity follows immediately. This shows that the near-nodal states play a crucial role in appearance of superconductivity. It also indicates that the "fermiology" approach in treating spin susceptibilities (references 32,41-47) might be at least as reasonable as the "stripe-ladder" picture from ref. 35.

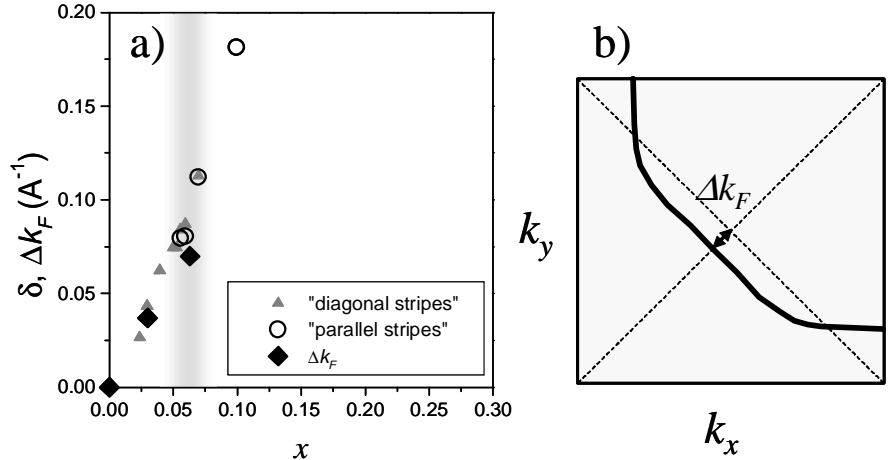

**Figure 5**: **a)** Doping dependence of the nodal $k_F$ in the "spin glass" regime of $La_{2-x}Sr_xO_4$ (solid diamonds) measured in ARPES.[6] Also shown is the incommensurability $\delta$ measured in neutron scattering close to the "spin glass" to superconducting transition (gray area) where the spin structure rotates from diagonal (x<0.06, solid triangles) to parallel (x>0.06, open circles) direction relative to the Cu-O bond. **b)** $k_F$ in a) is measured from $(\pi/2, \pi/2)$ point in the zone.

The role of near-nodal states in superconductivity might be only a secondary one: the one of phase coherence propagators, where due to the small superconducting gap $\Delta(k)$, these states have large coherence length $\xi(k) = v_F(k)/|\Delta(k)|$, and are able to stabilize the global superconducting phase[56] in the presence of inhomogeneities observed in STM[57,58,59] in the underdoped $Bi_2Sr_2CaCu_2O_{8+\delta}$. Also, such phase propagation carried by near-nodal states might be responsible for the "giant proximity effect," an observation where supercurrent runs through a thick layer of "normal metal" made out of an underdoped cuprate.[60] However, it might as well be that the superconductivity is a relatively weak order parameter, residing only in the near-nodal region, and is disrupted by some much larger order parameter with similar symmetry that dominates the rest of the Fermi surface in the underdoped regime.[61,62,63]

## 3. CONCLUSIONS

In conclusion, I have pointed out that some of less appreciated experimental results on cuprate superconductors indicate a very strong connection between dynamic spin susceptibility measured in inelastic neutron scattering and single-particle properties measured in ARPES and superconductivity itself. I have show that if dynamic spin susceptibility is used to calculate single-particle self energies, they agree in many aspects with the measured ones in ARPES. The interplay of spin fluctuations and superconductivity is particularly evident at the onset ($x\approx0.05$) and at the end ($x\approx0.3$) of superconducting "dome". At the onset, the static "diagonal" incommensurate spin density wave gives away to superconductivity, while at the "end" a simultaneous disappearance of spin fluctuations, of a coupling observed in ARPES ("kink") and of superconductivity shows that all three phenomena are inseparable.

## ACKNOWLEDGMENTS

I would like to acknowledge useful discussions with John Tranquada, Peter Johnson, Sasa Dordevic, Myron Strongin, Chris Homes, Alexei Tsvelik, Steve Kivelson and Doug Scalapino. The program was supported by the US DOE under contract number DE-AC02-98CH10886.